\documentclass[graybox,natbib]{svmult}

\usepackage{mathptmx}       
\usepackage{helvet}         
\usepackage{courier}        
\usepackage{type1cm}        
\usepackage{makeidx}         
\usepackage{graphicx}        
\usepackage{multicol}        
\usepackage[bottom]{footmisc}
\usepackage{mathrsfs}
\usepackage{bm}

\makeindex             

\begin{document}

\title*{An Introduction to Data Analysis in Asteroseismology}
\author{Tiago L.~Campante}
\institute{Tiago L.~Campante \at School of Physics and Astronomy, University of Birmingham, Birmingham B15 2TT, UK,\\
\email{campante@bison.ph.bham.ac.uk}\\ \\
Stellar Astrophysics Centre (SAC), Department of Physics and Astronomy, Aarhus University, Ny Munkegade 120, DK-8000 Aarhus C, Denmark\\ \\
Institut f\"{u}r Astrophysik, Georg-August-Universit\"{a}t G\"{o}ttingen, Friedrich-Hund-Platz 1, 37077 G\"{o}ttingen, Germany
}
%
%
\maketitle

\abstract{A practical guide is presented to some of the main data analysis concepts and techniques employed contemporarily in the asteroseismic study of stars exhibiting solar-like oscillations. The subjects of digital signal processing and spectral analysis are introduced first. These concern the acquisition of continuous physical signals to be subsequently digitally analyzed. A number of specific concepts and techniques relevant to asteroseismology are then presented as we follow the typical workflow of the data analysis process, namely, the extraction of global asteroseismic parameters and individual mode parameters (also known as peak-bagging) from the oscillation spectrum.
}

\section{Introduction}\label{sec:intro}
\textit{Solar-like oscillations} are excited by turbulent convection in the outer layers of stars \citep[see, e.g.,][and references therein]{JCD04}. Consequently, all stars cool enough to harbor an outer convective envelope may be expected to exhibit solar-like oscillations. Among several other classes of pulsating stars, solar-like oscillations are detectable in main-sequence core, and post-main-sequence shell, hydrogen-burning stars residing on the cool side of the Cepheid instability strip. The NASA \textit{Kepler} mission \citep{Kepler} has led to a revolution in the field of cool-star asteroseismology by allowing the detection of solar-like oscillations in several hundred solar-type stars (i.e., low-mass, main-sequence stars and cool subgiants) and in over ten thousand red giants \citep[for a review, see][]{ChaplinMiglioReview}. Of all these stars displaying solar-like oscillations about one hundred are \textit{Kepler} Objects of Interest (KOIs), i.e., candidate exoplanet-host stars \citep[e.g.,][]{Kepler444,Lundkvist}.

The present chapter is intended as a practical guide to some of the main data analysis concepts and techniques employed contemporarily in the asteroseismic study of stars exhibiting solar-like oscillations. The contents of this chapter strongly reflect the author's own experience as a data analyst. For that reason, special care has been taken to provide references to the work conducted by others, so that the reader can easily expand on the material presented herein.

\begin{figure}[t]
\centering
\includegraphics[scale=.35]{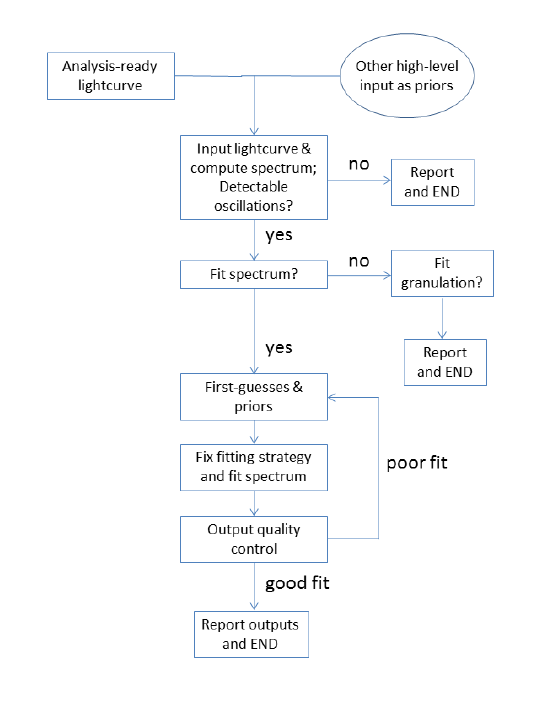}
\caption{Typical workflow of the data analysis process.}
\label{fig:workflow}       
\end{figure}

Pre-processing of light curves, although an integrant part of the data analysis process, is beyond the scope of this contribution. I therefore start by introducing the subjects of digital signal processing and spectral analysis in Sect.~\ref{sec:basics}. These concern the acquisition of continuous physical signals to be subsequently digitally analyzed. A number of specific concepts and techniques relevant to asteroseismology are then presented as we follow the typical workflow of the data analysis process (see Fig.~\ref{fig:workflow}). One must first establish whether signatures of solar-like oscillations are detectable in the power spectrum of the light curve. If they are, an attempt is made at extracting \textit{global asteroseismic parameters} from the data (Sect.~\ref{sec:globalpar}). One then establishes whether the oscillation spectrum is of sufficient quality to allow extraction of individual frequencies. If the answer is yes, \textit{individual mode parameters} are then extracted by fitting a multiparameter model to the oscillation spectrum, i.e., by \textit{peak-bagging} the oscillation spectrum (Sect.~\ref{sec:pbagging}).

\section{Digital signal processing and spectral analysis}\label{sec:basics}
Whereas some temporal phenomena can be understood through models in the time domain involving deterministic trends and/or stochastic autoregressive behavior, others are dominated by periodic behavior that is most effectively modeled in the frequency domain. The functional form of solar-like oscillations is that of a \textit{stochastically-excited harmonic oscillator}. This being a periodic functional form, the Fourier transform becomes the obvious choice for performing data analysis.

\subsection{Nyquist sampling theorem and aliasing}\label{sec:nyquist}
Let us consider the idealized case of a continuous signal $x(t)$ sampled by a set of impulse functions regularly spaced by $\Delta t$. Since the Fourier transform of such a set of impulse functions is another set of impulse functions with separation $1/\Delta t$ in the frequency domain, one can use the convolution theorem to show that the transform of the sampled signal is periodic:
\begin{equation}
\label{eq:sampling}
x(t) \sum_{n=-\infty}^{+\infty} \delta\left(t-n\,\Delta t\right) \Longleftrightarrow X(\nu) \ast \frac{1}{\Delta t} \sum_{n=-\infty}^{+\infty} \delta\left(\nu-\frac{n}{\Delta t}\right) \, ,
\end{equation}
where $X(\nu)$ is the Fourier transform of $x(t)$, the symbol `$\Longleftrightarrow$' indicates a Fourier pair and the symbol `$\ast$' denotes convolution.

The \textit{Nyquist sampling theorem} \citep{Nyquist,Shannon} states that if the Fourier transform of a continuous signal is band-limited, i.e., is zero for all $|\nu|\!\ge\!\nu_{\rm{lim}}$, then $x(t)$ can be uniquely reconstructed from a knowledge of its sampled values at uniform intervals of $\Delta t\!\leq\!1/(2\,\nu_{\rm{lim}})$. For a given uniform sampling interval $\Delta t$, the \textit{Nyquist frequency} is defined as $\nu_{\rm{Nyq}}\!=\!1/(2\Delta t)$. In case the continuous signal being sampled contains frequency components above the Nyquist frequency, these will give rise to an effect known as \textit{aliasing}, whereby the transform of the continuous signal is distorted due to spectral leakage. The signal is then said to be undersampled and can no longer be uniquely recovered.

The Nyquist frequency can be thought of as the highest useful frequency to search for in the power spectrum. However, based on astrophysical arguments, one can also accept frequencies above $\nu_{\rm{Nyq}}$ \citep{MurphySuperNyquist,ChaplinSuperNyquist}. Prospects for detecting solar-like oscillations in the \textit{super-Nyquist} regime of \textit{Kepler} long-cadence data, i.e., above the associated Nyquist frequency of $\sim\!283\:{\rm \mu Hz}$, are now being explored \citep{Yu16}. Targets of interest are cool subgiants and stars lying at the base of the red-giant branch.

Regular gaps in the light curve due to diurnal interruptions and, for data sets spanning more than a year, caused by the annual motion of the Earth, are usually present in observations carried out from the ground, giving rise also to frequency aliasing. \textit{Daily aliases}, appearing at splittings of $\pm1\:{\rm{cycle/day}}$ (or, equivalently, $\pm11.57\:{\rm{\mu Hz}}$), are particularly problematic when observing solar-like oscillations, since frequency separations of that same magnitude are common \citep[e.g.,][]{Procyon,Procyon2}.

\subsection{Filtering}\label{sec:timefilter}
Asteroseismic time series are often affected by low-frequency drifts, which can be either of instrumental origin or else intrinsic to the star. These low-frequency drifts introduce a background in the Fourier domain that ultimately leads to a degradation of the signal-to-noise ratio (SNR) in the oscillation modes. High-pass filters are widely used to reduce this effect while preserving the relevant signals.

Let us start by shedding some light on the process of \textit{smoothing} of a time series. Smoothing consists in convolving a signal $x(t)$ with a weighting function $w(t)$:
\begin{equation}
\label{eq:timefilter}
x_{\rm{low}}(t) = x(t) \ast w(t) \Longleftrightarrow X_{\rm{low}}(\nu) = X(\nu) \, W(\nu) \, ,
\end{equation}
where $X(\nu)$ and $W(\nu)$ are the transforms of $x(t)$ and $w(t)$, respectively. Conversely, a \textit{high-pass filter} can be implemented by simply computing $x_{\rm{high}}(t)\!=\!x(t)-x_{\rm{low}}(t)$:
\begin{equation}
\label{eq:timefilter2}
x_{\rm{high}}(t) \Longleftrightarrow X_{\rm{high}}(\nu) = X(\nu) \left[1-W(\nu)\right] \, .
\end{equation}

A commonly used high-pass filter in helioseismology is the backwards-difference filter \citep[][]{bdfilter}:
\begin{equation}
\label{eq:bdfilter}
x_{\rm{bd}}(t)=x(t)-x(t-t_0)=x(t)-\left[x(t)\ast\delta(t-t_0)\right] \, ,
\end{equation}
where a time shift $t_0$ has been considered. It becomes immediately obvious that $w(t)\!=\!\delta(t-t_0)$ in Eq.~(\ref{eq:timefilter}). Using Eq.~(\ref{eq:timefilter2}), one can then determine the transfer function of the backwards-difference filter:
\begin{equation}
\label{eq:transfer}
|1-W(\nu)|^2=\left[2 \sin\left(\frac{\pi}{2}\frac{\nu}{\nu_{\rm{c}}}\right)\right]^2 \, ,
\end{equation}  
where the cut-off frequency, $\nu_{\rm{c}}\!=\!1/(2\,t_0)$, has been introduced.

Typical examples of the weighting function $w(t)$ are a boxcar function, a triangular function (equivalent to the convolution of two boxcar functions) and a bell-shaped function (equivalent to the convolution of four boxcar functions or two triangular functions). The transform of the boxcar function is the sinc function and thus leads to an excessive ringing (or Gibbs-like) effect in the Fourier domain. Multiple-boxcar smoothing is therefore advisable.

\subsection{Power spectral density estimation}\label{sec:psdestimation}
Attention is first drawn to the estimation of the Fourier transform of $x(t)$ based on a finite number of samples. Suppose there are $N$ evenly spaced samples $x(t_n)\!=\!x(n\Delta t)$, with $n\!=\!0,1,\ldots,N\!-\!1$. The \textit{Discrete Fourier Transform}\footnote{\citet{FFT} have introduced the \textit{Fast Fourier Transform} (FFT), an efficient method of implementing the DFT.} (DFT) is defined as:
\begin{equation}
\label{eq:DFT}
X_{\rm{DFT}}(\nu_p)=\sum_{n=0}^{N-1} x(t_n)\,{\rm{e}}^{{\rm{i}}\,2\pi\nu_p t_n} \;\; {\rm{for}} \; \nu_p=p/(N\Delta t) \, , \; p=0,1,\ldots,N-1 \, .
\end{equation}
$X_{\rm{DFT}}(\nu_p)$ is the truncated transform of the sampled signal, which has periodicity $1/\Delta t$ or twice the Nyquist frequency. Then $p\!=\!0$ corresponds to the transform at zero frequency and $p\!=\!N/2$ to the value at $\pm\nu_{\rm{Nyq}}$. Values of $p$ between $N/2\!+\!1$ and $N\!-\!1$ correspond to the transform for negative frequencies.

Finally, I introduce the one-sided \textit{power density spectrum} or \textit{power spectrum}, $P(\nu_q)$, defined only for nonnegative frequencies (with $q\!=\!0,1,\ldots,N/2$):
\begin{eqnarray}
\label{eq:psd}
P(\nu_0) & = & \frac{\Delta t}{N} \left|X_{\rm{DFT}}(\nu_0)\right|^2 \, , \nonumber \\\nonumber \\
P(\nu_q) & = & \frac{\Delta t}{N} \left[\left|X_{\rm{DFT}}(\nu_p)\right|^2 + \left|X_{\rm{DFT}}(\nu_{N-p})\right|^2\right] \, , \\\nonumber \\
P(\nu_{N/2}) & = & \frac{\Delta t}{N} \left|X_{\rm{DFT}}(\nu_{N/2})\right|^2 \, , \nonumber
\end{eqnarray}
where $\nu_{N/2}\!=\!1/(2\Delta t)$ (i.e., the Nyquist frequency). Based on \textit{Parseval's theorem} \citep{Parseval}, we may then normalize $P(\nu_q)$ according to
\begin{equation}
\sum_{q=0}^{N/2} P(\nu_q)\,\Delta\nu = \frac{1}{N} \sum_{n=0}^{N-1} x^2(t_n) \, .
\end{equation}

According to the \textit{Wiener--Khintchine theorem} \citep{Wiener,Khintchine}, the power spectrum and the autocorrelation function, $\phi(\tau)$, are a Fourier pair:
\begin{equation}
\label{WK}
\phi(\tau)=\int_{-\infty}^{+\infty} P(\nu)\,{\rm{e}}^{-{\rm{i}}\,2\pi\nu\tau}\,{\rm{d}}\nu \Longleftrightarrow P(\nu)=\int_{-\infty}^{+\infty} \phi(\tau)\,{\rm{e}}^{{\rm{i}}\,2\pi\nu\tau}\,{\rm{d}}\tau \, ,
\end{equation}  
where 
\begin{equation}
\phi(\tau)=\lim_{T\to\infty} \frac{1}{T} \int_{-T/2}^{T/2} x(t)x(t+\tau)\,{\rm{d}}t \, .
\end{equation}
The Wiener--Khintchine theorem is absolutely crucial to understanding the spectral analysis of random processes. It straightforwardly explains, for instance, why white noise, whose autocorrelation function is the Dirac delta function, has constant power spectral density.

\subsection{Power spectrum statistics and hypothesis testing}\label{sec:psdstats}
In the following I consider the \textit{statistics} of the power spectrum of a pure noise signal \citep[see also][]{AppourchauxCrash}. Let $x(t)$ represent a random process from which a finite number of samples $x(t_n)$ are drawn. The samples are assumed to be independent and identically distributed (i.i.d.), and the process is further assumed to be stationary, with ${\rm{E}}\left[x(t_n)\right]\!=\!0$ and ${\rm{E}}\left[x^2(t_n)\right]\!=\!\sigma_0^2$ for all $n$. The DFT of the set $x(t_n)$ may be decomposed into its real and imaginary parts as:
\begin{eqnarray}
\label{eq:stats}
X_{\rm{DFT}}(\nu_p) & = & X_{\rm{DFT}}^{\rm{Re}}(\nu_p) + {\rm{i}}\,X_{\rm{DFT}}^{\rm{Im}}(\nu_p) \nonumber \\
& = & \sum_{n=0}^{N-1} x(t_n)\cos(2\pi\nu_pt_n) + {\rm{i}}\,\sum_{n=0}^{N-1} x(t_n)\sin(2\pi\nu_pt_n) \, . 
\end{eqnarray}
It follows from the Central Limit theorem that, for large $N$, both $X_{\rm{DFT}}^{\rm{Re}}$ and $X_{\rm{DFT}}^{\rm{Im}}$ are normally distributed with
\begin{equation}
\label{eq:stats2}
{\rm{E}}\left[X_{\rm{DFT}}^{\rm{Re}}(\nu_p)\right]={\rm{E}}\left[X_{\rm{DFT}}^{\rm{Im}}(\nu_p)\right]=0 \, ,
\end{equation} 
\begin{equation}
\label{eq:stats3}
{\rm{E}}\left[\left(X_{\rm{DFT}}^{\rm{Re}}(\nu_p)\right)^2\right]={\rm{E}}\left[\left(X_{\rm{DFT}}^{\rm{Im}}(\nu_p)\right)^2\right]=\frac{N}{2}\sigma_0^2 \, .
\end{equation}
Finally, since $X_{\rm{DFT}}^{\rm{Re}}$ and $X_{\rm{DFT}}^{\rm{Im}}$ are independent and have the same normal distribution, the power spectrum, $\left|X_{\rm{DFT}}\right|^2$, then has by definition a \textit{chi-squared distribution with 2 degrees of freedom} (i.e., $\chi^2_2$).

Adopting $\left|X_{\rm{DFT}}\right|^2 \Delta t/N$ as our normalization of the power spectrum yields a constant power spectral density for the noise of $\sigma_0^2\Delta t$ and variance $(\sigma_0^2\Delta t)^2$. Consequently, as $N$ tends to infinity by sampling a longer stretch of data, the variance in the power spectrum remains unchanged. Furthermore, the probability density, $p(z)$, that the observed power spectrum takes a particular value $z$ at a fixed frequency bin is given by
\begin{equation}
\label{eq:stats4}
p(z)=\frac{1}{\langle z \rangle}\,\exp\left(-\frac{z}{\langle z \rangle}\right) \, ,
\end{equation} 
where $\langle z \rangle\!=\!\sigma_0^2\Delta t$. Equation (\ref{eq:stats4}) enables one to derive the probability that the power in one bin is greater than $m$ times the mean level of the continuum, $\langle z \rangle$:
\begin{equation}
\label{eq:hyptest1}
F(m)={\rm{e}}^{-m} \, .
\end{equation} 
For instance, a \textit{confidence level} of 99\% or, equivalently, a \textit{false alarm probability} of 1\%, leads to $m\!\approx\!4.6$. For a frequency band containing $M$ bins, the probability that at least one bin has a normalized power greater than $m$ is then:
\begin{equation}
\label{eq:hyptest2}
F_M(m)=1-(1-{\rm{e}}^{-m})^M \, ,
\end{equation}  
which approximates to $F_M(m)\!=\!M{\rm{e}}^{-m}$ for ${\rm{e}}^{-m}\!\ll\!1$.

In astrophysics it is very common to deal with unevenly sampled time series. In that event, an existing frequentist statistic known as the \textit{Lomb--Scargle periodogram}\footnote{Fast computation of the periodogram is achieved using the algorithm presented in \citet{LSP}, whose trick is to carry out extirpolation of the data onto a regular mesh and subsequently employ the FFT.} is widely used as an estimator of the power spectral density. The Lomb--Scargle periodogram can be formulated either as a modified Fourier analysis or as a least-squares regression of the data set to sine waves with a range of frequencies. It has the attractive property of retaining the $\chi^2_2$ statistics.

\subsection{Non-Fourier periodograms}\label{sec:nonFourier}
Astronomers have developed and extensively used a variety of \textit{non-Fourier periodograms} for period searches in unevenly spaced data sets \citep[e.g.,][]{Clarke02}. The most common strategy involves folding the data modulo a trial period, computing a statistic on the folded time series (now a function of phase rather than time), and plotting the statistic for all independent periods. These methods measure the strength of signals that are strictly periodic, but not necessarily sinusoidal in shape. They are also relatively insensitive to the duration and uneven spacing of the data set, and some methods readily permit heteroscedastic weighting from measurement errors. An overview of the application of non-Fourier periodograms to asteroseismic data sets is given in sect.~5.2 of the book by \citet{AsteroBook}.

\section{Extracting global asteroseismic parameters}\label{sec:globalpar}
In order to fully characterize a star using asteroseismology, it is desirable that we obtain precise estimates of individual mode parameters (e.g., frequencies, amplitudes and linewidths). However, this is only possible for data above a certain SNR. \textit{Global asteroseismic parameters}, indicative of the overall stellar structure, are on the other hand readily extractable using automated pipelines that are able to incorporate data with a lower SNR and for which a full peak-bagging analysis is not always possible. Furthermore, the automated nature of these pipelines is required if we are to efficiently exploit the large volumes of data made available by current and future space-based missions \citep[][]{PLATO,CampanteTESS}.

In this section I introduce an automated pipeline\footnote{A comparison of different pipelines used to extract global asteroseismic parameters is presented in \citet{Verner11}.} which has been originally designed to extract global asteroseismic parameters of main-sequence and subgiant stars from \textit{Kepler} power spectra \citep{Campante10,CampantePhD}. This pipeline allows extracting the following information from the power spectrum (points 1--4 are covered below):
\begin{enumerate}
\item Frequency range of the oscillations;
\item Parameterization of the stellar background signal;
\item Average large frequency separation, $\Delta\nu$;
\item Frequency of maximum amplitude, $\nu_{\rm{max}}$;
\item Maximum mode amplitude, $A_{\rm{max}}$.
\end{enumerate}

\subsection{Detectability of oscillations}\label{sec:detectability}
We want to look for a frequency range in the power spectrum in which peaks appear at nearly regular intervals, one of the main signatures of the presence of solar-like oscillations. I note that the assumption of \textit{quasi-regularity} may, however, be too strong in the case of evolved stars due to the presence of mixed modes. We start by partitioning the power spectrum into overlapping windows of variable width, $w$. The width $w$ depends on the central frequency of the window, $\nu_{\rm{central}}$, used as a proxy for $\nu_{\rm{max}}$. We make use of the fact that the width of the p-mode bump approximately scales with $\nu_{\rm{max}}$ \citep[e.g.,][]{Stello07,Mosser10}, and so $w$ is defined as $w\!=\!(\nu_{\rm{central}}/\nu_{{\rm{max}},\odot})w_\odot$.

The next step consists in computing the \textit{power spectrum of the power spectrum}, PS$\otimes$PS, for each of these frequency windows. The presence of prominent features in the PS$\otimes$PS around the predicted\footnote{The predicted value of $\Delta\nu$ is computed according to the relation $\Delta\nu\!\propto\!\nu_{\rm{central}}^{0.77}$ \citep{Stello09}.} values of $\Delta\nu/2$, $\Delta\nu/4$, and $\Delta\nu/6$ (the first, second, and third harmonics, respectively) is then examined. An hypothesis test is subsequently applied, whereby the presence of oscillations in a given window is established if the probability of the three above features being due to noise is less than 1\%. Finally, the frequency range of the oscillations is determined based on the overall span of the windows with detected oscillations.

\begin{figure}[t]
\centering
\includegraphics[scale=.3]{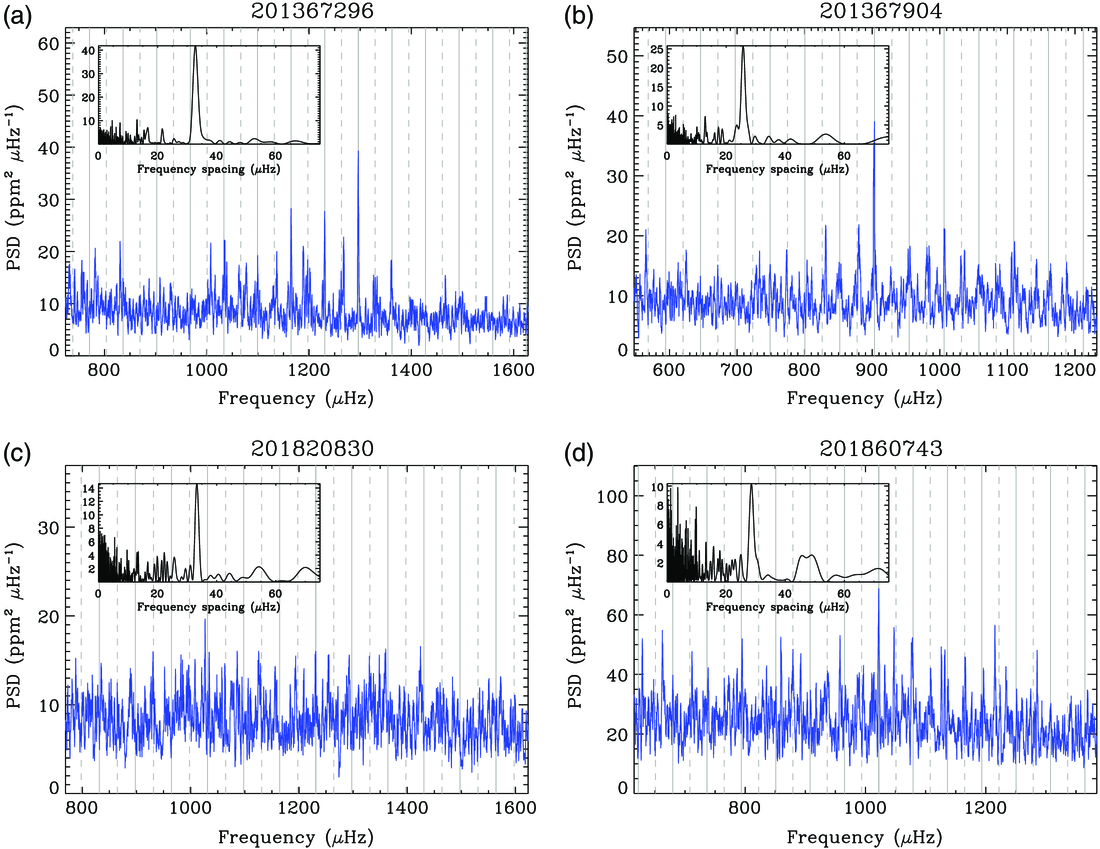}
\caption{K2 power spectra (slightly smoothed) of four solar-type stars with detected oscillations. Insets show the PS$\otimes$PS, computed from the region around $\nu_{\rm{max}}$. From \citet{Chaplin15}.}
\label{fig:detect}       
\end{figure}

Figure \ref{fig:detect} shows the detection of oscillations in the K2 power spectra of four solar-type stars. Sets of vertical gray solid and dashed lines are separated by the estimated $\Delta\nu$, and mark the spacing on which we would expect to see modes. The insets show the PS$\otimes$PS, computed from the region around $\nu_{\rm{max}}$. The significant peaks in the PS$\otimes$PS lie at $\Delta\nu/2$ and are a signature of the near-regular spacing of solar-like oscillations.

\subsection{Background signal}\label{sec:background}
The model of the \textit{stellar background signal} is kept simple, merely containing a granulation component and photon shot noise. We fit this model to a smoothed version of the power spectrum employing a nonlinear least-squares fitting algorithm. 

The frequency range of the oscillations (if detected) is excluded from the fitting window. The fitting window starts at $100\:{\rm{\mu Hz}}$ to allow for the decay of any possible activity component, characterized by considerably longer timescales, and extends all the way up to the Nyquist frequency of {\it Kepler} short-cadence data ($\sim\!8300\:{\rm{\mu Hz}}$).

The granulation component is represented by a \textit{Harvey-like profile} \citep[e.g.,][and references therein]{Kallinger14} to which an offset is added to account for the \textit{shot noise} component:
\begin{equation}
B(\nu) = B_0 + \eta^2(\nu) \left[\frac{B_{\rm gran}}{1+(2\pi\nu\,\tau_{\rm gran})^a}\right] \, ,
\end{equation}
where $B_{\rm gran}$ is the height at $\nu\!=\!0$ of the granulation component, $\tau_{\rm gran}$ is the characteristic turnover timescale and $a$ calibrates the amount of memory in the process. Such a functional form is representative of a random non-harmonic field whose autocorrelation decays exponentially with time. The attenuation factor $\eta^2(\nu)$ takes into account the apodization of the oscillation signal due to the finite integration time.

\begin{figure}[t]
\centering
\includegraphics[scale=.4]{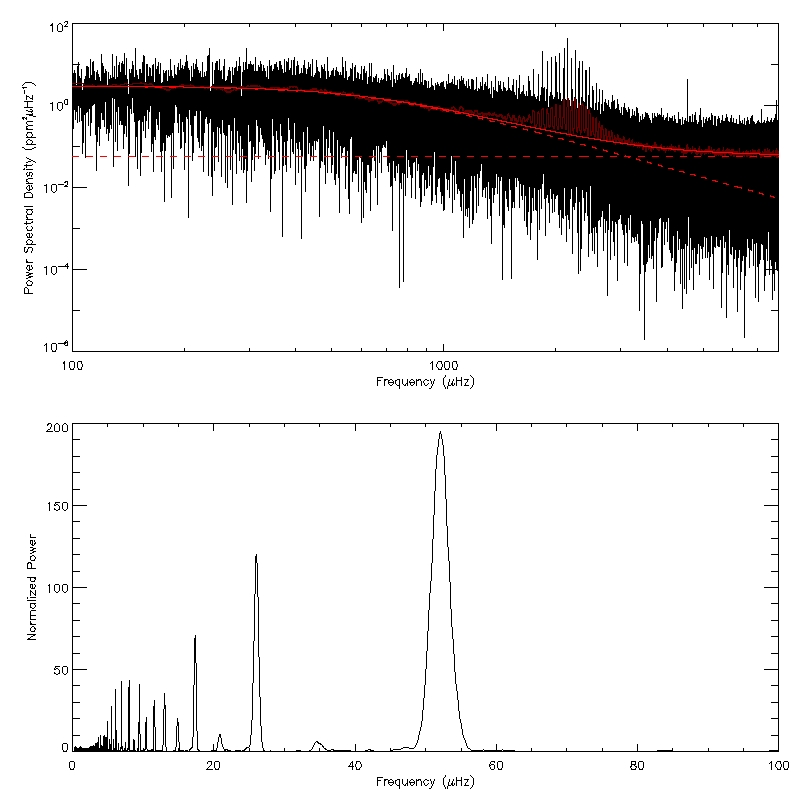}
\caption{Output from the analysis of the \textit{Kepler} light curve of the bright G-type dwarf 16 Cyg A. \textit{Top panel:} Modeling the stellar background signal. \textit{Bottom panel:} Detection of oscillations in the PS$\otimes$PS. From \citet{CampantePhD}.}
\label{fig:pipeline}       
\end{figure}

The top panel of Fig.~\ref{fig:pipeline} displays the smoothed power spectrum of 16 Cyg A (dark red) overlaid on the original power spectrum (black). The fit to the background signal (red solid line) and both its components (red dashed lines) are also shown. The bottom panel displays the PS$\otimes$PS over the frequency range of the oscillations. The features at $\Delta\nu/2$ ($\sim\!52\:{\rm{\mu Hz}}$), $\Delta\nu/4$ ($\sim\!26\:{\rm{\mu Hz}}$) and $\Delta\nu/6$ ($\sim\!17\:{\rm{\mu Hz}}$) are conspicuous.

\subsection{Large frequency separation ($\Delta\nu$)}\label{sec:Dnu}
In order to estimate the \textit{average large frequency separation}, $\Delta\nu$, we compute the PS$\otimes$PS over the frequency range of the oscillations. The feature at $\Delta\nu/2$ (first harmonic) in the PS$\otimes$PS is then located and its power-weighted centroid computed to provide an estimate of $\Delta\nu$. The standard deviation of grouped data, given by $\sqrt{\left[\sum hx^2 - (\sum hx)^2/\sum h\right]\,/\,(\sum h - 1)}$, is adopted as the error on $\Delta\nu$, meaning that the feature in the PS$\otimes$PS is interpreted as an assembly of spectral heights ($h$) over a number of bins (with midpoint $x$).

\subsection{Frequency of maximum amplitude ($\nu_{\rm{max}}$)}\label{sec:nu_max}
In order to estimate the \textit{frequency of maximum amplitude}, $\nu_{\rm{max}}$, we average the p-mode power (after subtraction of the background fit) over contiguous rectangular windows of width $2\Delta\nu$ and convert to power per radial mode by multiplying by $\Delta\nu/c$, where $c$ measures the effective number of modes per order \citep[see][]{Kjeldsen08}. An estimate of $\nu_{\rm{max}}$ is then given by the power-weighted centroid, with the associated uncertainty derived from the standard deviation of grouped data (see Sect.~\ref{sec:Dnu}).

\section{Peak-bagging}\label{sec:pbagging}
In this section I introduce a Bayesian peak-bagging tool that employs \textit{Markov chain Monte Carlo} (MCMC) techniques \citep[e.g.,][]{Campante11,HandCamp,CampantePhD,Campante16}. Besides making it possible to incorporate relevant prior information through Bayes' theorem, this tool also allows obtaining the marginal probability density function (pdf) for each of the model parameters. Such techniques are in many ways an extension of the \textit{Maximum Likelihood Estimation} (MLE) methods originally introduced in helioseismology \citep[][]{DuvallHarvey86,Anderson90}.

\subsection{Power spectrum of a solar-like oscillator}\label{sec:ps_solarlike}
Understanding the characteristics of the power spectrum of a solar-like oscillator is fundamental in order to extract information on the physics of the modes. The stochastic driving of a damped oscillator can be described by
\begin{equation}
\label{eq:oscillator}
\frac{{\rm{d}}^2}{{\rm{d}}t^2}y(t) + 2\eta\,\frac{{\rm{d}}}{{\rm{d}}t}y(t) + \omega_0^2\,y(t)=f(t) \, ,
\end{equation}
where $y(t)$ is the amplitude of the oscillator, $\eta$ is the linear damping rate, $\omega_0$ is the frequency of the undamped oscillator and $f(t)$ is a random forcing function. The Fourier transform of Eq.~(\ref{eq:oscillator}) is then expressed as
\begin{equation}
\label{eq:oscillator2}
-\omega^2\,Y(\omega)-{\rm{i}}\,2\eta\omega\,Y(\omega) + \omega_0^2\,Y(\omega)=F(\omega).
\end{equation}

When a realization of $y(t)$ is observed for a finite amount of time, an estimate of the power spectrum is then given by
\begin{equation}
\label{eq:oscillator3}
P(\omega)=|Y(\omega)|^2=\frac{|F(\omega)|^2}{(\omega_0^2-\omega^2)^2 + 4\,\eta^2\omega^2} \, .
\end{equation}
In the limit of taking the average of an infinite number of realizations, and assuming the damping rate to be very small compared to the frequency of oscillation, one obtains near the resonance the following expression for the \textit{limit spectrum}:
\begin{equation}
\label{eq:lorentzprof}
\langle P(\omega)\rangle \simeq \frac{1}{4\,\omega_0^2} \frac{\langle P_f(\omega)\rangle}{(\omega-\omega_0)^2+\eta^2} \, .
\end{equation}  
The average power spectrum of the random forcing function, $\langle P_f(\omega)\rangle$, is a slowly-varying function of frequency. The result is thus a \textit{Lorentzian profile}, characterized by the central frequency $\omega_0$ and a width determined by the linear damping rate $\eta$.

\begin{figure}[t]
\centering
\includegraphics[scale=.4]{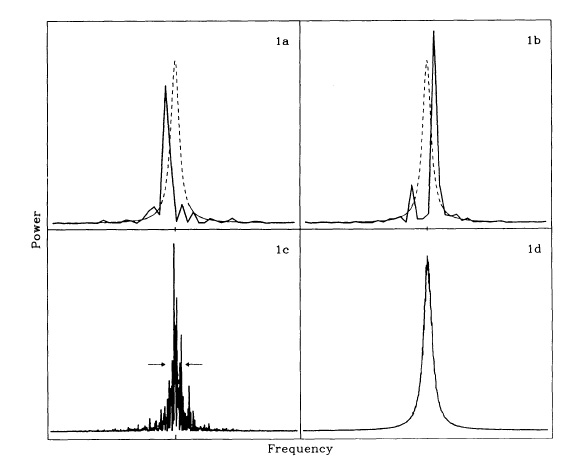}
\caption{Lorentzian profile (limit spectrum) and the erratic behavior of the power spectrum. From \citet{Anderson90}.}
\label{fig:lorentz}       
\end{figure}

Panels (a) and (b) in Fig.~\ref{fig:lorentz} display two realizations of the same limit spectrum. Both power spectra appear as an erratic function concealing the underlying Lorentzian profile. Panel (c) displays a realization of the same limit spectrum, although with a resolution twenty times higher. Increasing the total observational span, hence the resolution, did nothing to reduce the variance in the power spectrum (cf.~Sect.~\ref{sec:psdstats}). Panel (d) displays the average of a large number of realizations with the same resolution as in (c), thus converging to the limit spectrum.

\subsection{Modeling the power spectrum}\label{sec:ps_model}
\begin{figure}[t]
\centering
\includegraphics[scale=.65]{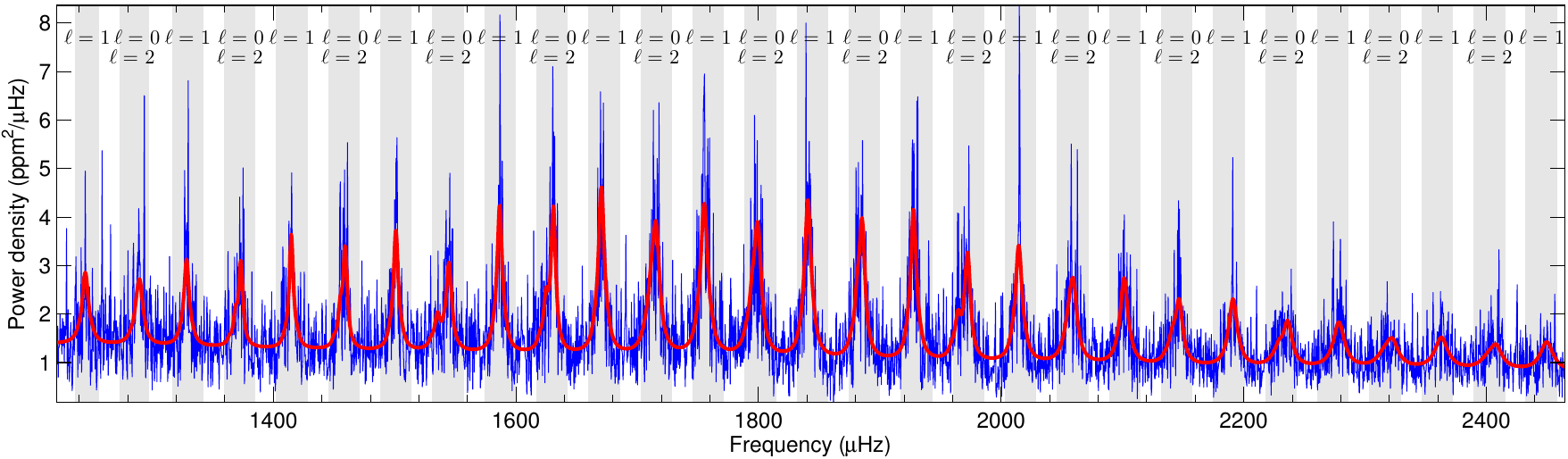}
\caption{Power spectrum of the F-type star HD~49933 (blue) based on 180 days of \textit{CoRoT} photometry and best-fitting model (red). Shaded areas indicate the ranges of the uniform priors on the mode frequencies. From \citet{HandCamp}.}
\label{fig:ps_model}       
\end{figure}

We are primarily interested in performing a \textit{global fit} to the power spectrum, whereby the observed modes are fitted simultaneously over a broad frequency range. We thus model the limit oscillation spectrum as a sum of standard Lorentzian profiles, $O(\nu)$, which sit atop a background signal described by $B(\nu)$:
\begin{eqnarray}
\label{eq:specmodel}
P(\nu;\bm{\lambda}) & = & O(\nu) + B(\nu) \nonumber \\ 
& = & \sum_{n',l} \sum_{m=-l}^{l} \frac{\mathscr{E}_{l m}(i_{\rm s}) H_{n'l}}{1+\left[\frac{2(\nu - \nu_{n'l0} - m\nu_{\rm s})}{\Gamma_{n'lm}}\right]^2} + B(\nu) \, ,
\end{eqnarray}
where $\bm{\lambda}$ represents the set of model parameters. The inner sum in the above equation runs over the azimuthal components $\{m\}$ of each multiplet $\{n',l\}$, while the outer sum runs over the selection of observed modes. Figure \ref{fig:ps_model} shows the power spectrum of HD~49933 (blue) based on 180 days of \textit{CoRoT} photometry. The best-fitting model (red) is overlaid, with the shaded areas indicating the ranges of the uniform priors (see Sect.~\ref{sec:parestimation}) on the mode frequencies.

At a given frequency bin $j$, the probability density, $f(P_j;\bm{\lambda})$, that the observed power spectrum takes a particular value $P_j$ is related to the limit spectrum, $P(\nu_j;\bm{\lambda})$, by (cf.~Eq.~\ref{eq:stats4})
\begin{equation}
\label{eq:pdfspec}
f(P_j;\bm{\lambda})=\frac{1}{P(\nu_j;\bm{\lambda})} \exp\left[-\frac{P_j}{P(\nu_j;\bm{\lambda})}\right] \, .
\end{equation}   
We now want to specify the \textit{likelihood function}, i.e., the joint pdf of the data sample $\{P_j\}$. Assuming the frequency bins to be uncorrelated, the joint pdf is simply given by the product of $f(P_j;{\boldsymbol\lambda})$ over some frequency interval of interest spanned by $j$:
\begin{equation}
\label{eq:eqlikelihood}
L(\bm{\lambda})=\prod_{j} f(P_j;\bm{\lambda}) \, .
\end{equation}

\subsection{Bayesian parameter estimation using MCMC}\label{sec:bayesian}
I now describe the formalism of a Bayesian approach to \textit{parameter estimation} and \textit{model comparison} that employs an MCMC algorithm. Let us consider a set of competing hypotheses, $\{H_i\}$, assumed to be mutually exclusive. One should be able to assign a probability, $p(H_i|D,I)$, to each hypothesis taking into account the observed data, $D$, and any available prior information, $I$. This is done through \textit{Bayes' theorem} \citep{Bayes}:
\begin{equation}
p(H_i|D,I)=\frac{p(H_i|I)p(D|H_i,I)}{p(D|I)} \, .
\label{eq:Bayes1}
\end{equation}
The probability of the hypothesis $H_i$ in the absence of $D$ is called the \textit{prior probability}, $p(H_i|I)$, whereas the probability including $D$ is called the \textit{posterior probability}, $p(H_i|D,I)$. The quantity $p(D|H_i,I)$ is called the \textit{likelihood} of $H_i$. The denominator $p(D|I)$ is the \textit{global likelihood} for the entire class of hypotheses. The sum of the posterior probabilities over the hypothesis space of interest is unity, hence one has:
\begin{equation}
p(D|I)=\sum_i p(H_i|I)p(D|H_i,I) \, .
\label{eq:globlike}
\end{equation}  

\subsubsection{Parameter estimation}\label{sec:parestimation}
If a particular hypothesis, i.e., a given model $M$ describing the physical process, is assumed true, then the hypothesis space of interest concerns the values taken by the model parameters, $\bm{\lambda}$. These parameters are continuous and one will be interested in obtaining their pdf. The global likelihood of model $M$ is then given by the continuous counterpart of Eq.~(\ref{eq:globlike}):
\begin{equation}
p(D|I)=\int p(\bm{\lambda}|I)p(D|\bm{\lambda},I) d\bm{\lambda} \, .
\label{eq:globlike2}
\end{equation}
We restate Bayes' theorem to account for this new formalism:
\begin{equation}
p(\bm{\lambda}|D,I)=\frac{p(\bm{\lambda}|I)p(D|\bm{\lambda},I)}{p(D|I)} \, ,
\label{eq:Bayes2}
\end{equation}
where $p(D|I)$ plays the role of a normalization constant. Ultimately, we are interested in using MCMC techniques to map the posterior pdf, $p(\bm{\lambda}|D,I)$. The procedure of \textit{marginalization} allows computation of the posterior pdf for a subset of parameters $\bm{\lambda}_A$ by integrating over the remaining parameters (or \textit{nuisance parameters}) $\bm{\lambda}_B$:
\begin{equation}
p(\bm{\lambda}_A|D,I)=\int p(\bm{\lambda}_A,\bm{\lambda}_B|D,I) d\bm{\lambda}_B \, .
\end{equation}

\subsubsection{Model comparison}\label{sec:modelcomp}
The problem of model comparison is analogous to that of parameter estimation. When facing a situation in which several parameterized models are available for describing the same physical process, one expects Bayes' theorem to allow for a statistical comparison between such models. Bayesian model comparison has a built-in \textit{Occam's razor} by which a complex model is automatically penalized, unless the available data justify its additional complexity. Competing models may be either intrinsically different models or else similar but with varying number of parameters (i.e., nested models), or even the same model with different priors affecting its parameters.

Given two or more competing models and our prior information, $I$, being in the present context that one and only one of the models is true, we can assign individual probabilities similarly to what has been done in Eq.~(\ref{eq:Bayes1}), after replacing $H_i$ by $M_i$:
\begin{equation}
p(M_i|D,I)=\frac{p(M_i|I)p(D|M_i,I)}{p(D|I)} \, ,
\label{eq:Bayes3}
\end{equation}
where the global likelihood of model $M_i$, $p(D|M_i,I)$, also called the \textit{evidence} of the model, is given by Eq.~(\ref{eq:globlike2}). We are often interested in computing the ratio of the probabilities of two competing models:
\begin{equation}
O_{ij}\equiv\frac{p(M_i|D,I)}{p(M_j|D,I)}=\frac{p(M_i|I)p(D|M_i,I)}{p(M_j|I)p(D|M_j,I)}=\frac{p(M_i|I)}{p(M_j|I)}B_{ij} \, ,
\end{equation}
where $O_{ij}$ is the \textit{odds ratio} in favor of model $M_i$ over model $M_j$, $B_{ij}$ is the so-called \textit{Bayes' factor} and $p(M_i|I)/p(M_j|I)$ is the \textit{prior odds ratio}. The Bayesian odds ratio is the product of the ratio of the prior probabilities of the models and the ratio of their global likelihoods.

\subsubsection{Markov chain Monte Carlo}\label{sec:mcmc}
The need becomes clear for a mathematical tool that is able to efficiently evaluate the multidimensional integrals required in the computation of the marginal distributions. The aim is to draw samples from the \textit{target distribution}, $p(\bm{\lambda}|D,I)$, by constructing a pseudo-random walk in parameter space such that the number of samples drawn from a particular region is proportional to its posterior density. This is achieved by generating a \textit{Markov chain}, whereby a new sample, $\bm{\lambda}_{t+1}$, depends on the previous sample, $\bm{\lambda}_{t}$, according to a time-independent quantity called the \textit{transition kernel}, $p(\bm{\lambda}_{t+1}|\bm{\lambda}_t)$. After a burn-in phase, $p(\bm{\lambda}_{t+1}|\bm{\lambda}_t)$ should be able to generate samples of $\bm{\lambda}$ with a probability density converging on the target distribution.

\begin{figure}[t]
\centering
\includegraphics[scale=1.0]{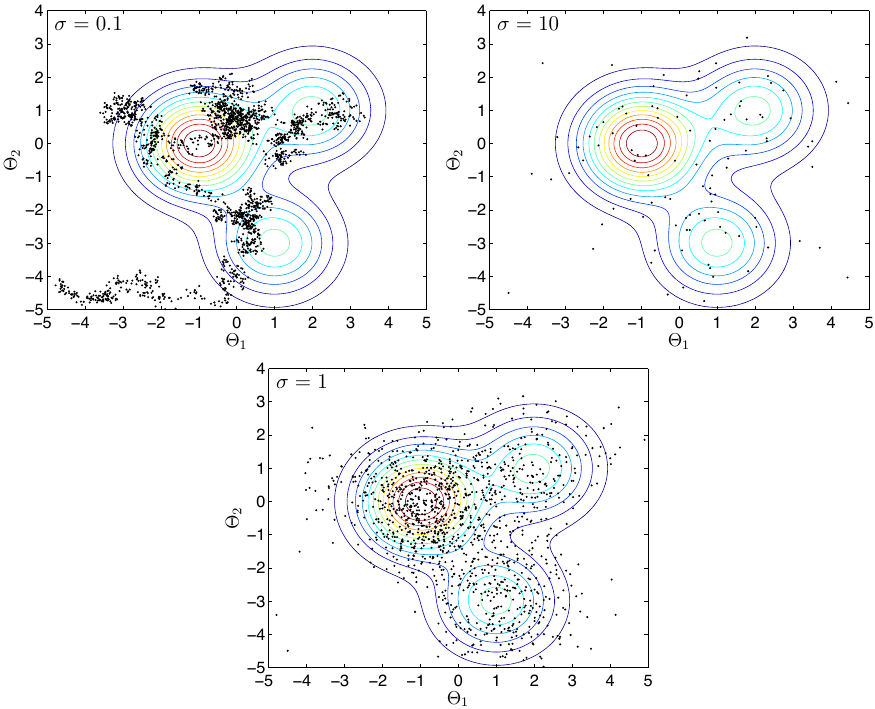}
\caption{Two-dimensional MCMC simulations of a triple-peaked posterior. The same target distribution is sampled by three chains, each characterized by a different set $\{\sigma\}$ of step sizes in parameter space. From \citet{HandCamp}.}
\label{fig:mcmc}       
\end{figure}

We generate a Markov chain by using the \textit{Metropolis--Hastings algorithm} \citep{Metropolis,Hastings}. Let us denote the current sample by $\bm{\lambda}_t$. We would like to steer the Markov chain toward the next sampling state, $\bm{\lambda}_{t+1}$, by first proposing a new sample, $\bm{\xi}$, to be drawn from a proposal distribution, $q(\bm{\xi}|\bm{\lambda}_t)$, which can have almost any form. The proposed sample is then accepted with a probability given by:
\begin{equation}
\alpha(\bm{\lambda}_t,\bm{\xi})=\min(1,r)=\min\left[1,\frac{p(\bm{\xi}|D,I)}{p(\bm{\lambda}_t|D,I)}\frac{q(\bm{\lambda}_t|\bm{\xi})}{q(\bm{\xi}|\bm{\lambda}_t)}\right] \, ,
\end{equation}
where $\alpha(\bm{\lambda}_t,\bm{\xi})$ is the \textit{acceptance probability} and $r$ is called the \textit{Metropolis ratio}. If $\bm{\xi}$ is not accepted, then the chain will keep the current sampling state, i.e., $\bm{\lambda}_{t+1}\!=\!\bm{\lambda}_t$. Figure \ref{fig:mcmc} shows the output from three two-dimensional MCMC simulations of the same triple-peaked posterior.

Once the posterior pdf, $p(\bm{\lambda}|D,I)$, has been mapped, the procedure of marginalization becomes trivial. The marginal posterior distribution of a given parameter $\lambda$, $p(\lambda|D,I)$, is then simply obtained by collecting its samples in a normalized histogram. An estimate of the $k$-th moment of $\lambda$ about the origin is then given by
\begin{equation}
\langle\lambda^k\rangle\equiv\int \lambda^k p(\lambda|D,I) d\lambda \approx \frac{1}{N}\sum\lambda_t^k \, ,
\end{equation}
where $N$ is the total number of samples.

The basic Metropolis--Hastings algorithm runs the risk of becoming stuck in a local mode of the target distribution. A way of overcoming this is to employ \textit{parallel tempering}, whereby a discrete set of progressively flatter versions of the target distribution is created by introducing a \textit{tempering parameter}, $\gamma$. We modify Eq.~(\ref{eq:Bayes2}) to generate the tempered distributions:
\begin{equation}
p(\bm{\lambda}|D,\gamma,I) \propto p(\bm{\lambda}|I) p(D|\bm{\lambda},I)^\gamma \, , \;\;\; 0<\gamma\leq1 \, .
\end{equation}
For $\gamma\!=\!1$, we retrieve the target distribution, while distributions with $\gamma\!<\!1$ are effectively flatter versions of the target distribution. By running such a set of chains in parallel and allowing their parameter states to swap, we increase the mixing properties of the Markov chain. 

Furthermore, the Metropolis--Hastings algorithm can be refined by implementing a statistical control system \citep[e.g.,][]{GregoryBook} allowing to automatically fine-tune the proposal distribution during the burn-in phase (see Fig.~\ref{fig:mcmc}).

\begin{acknowledgement}
The author acknowledges the support of the UK Science and Technology Facilities Council (STFC). Funding for the Stellar Astrophysics Centre is provided by The Danish National Research Foundation (Grant DNRF106).
\end{acknowledgement}

\bibliographystyle{apj}
\bibliography{biblio}

\begin{thebibliography}{}
\expandafter\ifx\csname natexlab\endcsname\relax\def\natexlab#1{#1}\fi

\bibitem[{{Aerts} {et~al.}(2010){Aerts}, {Christensen-Dalsgaard}, \&
  {Kurtz}}]{AsteroBook}
{Aerts}, C., {Christensen-Dalsgaard}, J., \& {Kurtz}, D.~W. 2010,
  {Asteroseismology} (Springer)

\bibitem[{{Anderson} {et~al.}(1990){Anderson}, {Duvall}, \&
  {Jefferies}}]{Anderson90}
{Anderson}, E.~R., {Duvall}, Jr., T.~L., \& {Jefferies}, S.~M. 1990, ApJ, 364,
  699

\bibitem[{{Appourchaux}(2013)}]{AppourchauxCrash}
{Appourchaux}, T. 2013, in Asteroseismology, ed. P.~L. {Pall{\'e}} \&
  C.~{Esteban}, Canary Islands Winter School of Astrophysics (Cambridge
  University Press), 123

\bibitem[{{Arentoft} {et~al.}(2008){Arentoft}, {Kjeldsen}, {Bedding}, {Bazot},
  {Christensen-Dalsgaard}, {Dall}, {Karoff}, {Carrier}, {Eggenberger},
  {Sosnowska}, {Wittenmyer}, {Endl}, {Metcalfe}, {Hekker}, {Reffert}, {Butler},
  {Bruntt}, {Kiss}, {O'Toole}, {Kambe}, {Ando}, {Izumiura}, {Sato}, {Hartmann},
  {Hatzes}, {Bouchy}, {Mosser}, {Appourchaux}, {Barban}, {Berthomieu},
  {Garcia}, {Michel}, {Provost}, {Turck-Chi{\`e}ze}, {Marti{\'c}}, {Lebrun},
  {Schmitt}, {Bertaux}, {Bonanno}, {Benatti}, {Claudi}, {Cosentino}, {Leccia},
  {Frandsen}, {Brogaard}, {Glowienka}, {Grundahl}, \& {Stempels}}]{Procyon}
{Arentoft}, T., {Kjeldsen}, H., {Bedding}, T.~R., {et~al.} 2008, ApJ, 687, 1180

\bibitem[{{Bayes} \& {Price}(1763)}]{Bayes}
{Bayes}, T., \& {Price}, R. 1763, Philosophical Transactions, 53, 370

\bibitem[{{Bedding} {et~al.}(2010){Bedding}, {Kjeldsen}, {Campante},
  {Appourchaux}, {Bonanno}, {Chaplin}, {Garcia}, {Marti{\'c}}, {Mosser},
  {Butler}, {Bruntt}, {Kiss}, {O'Toole}, {Kambe}, {Ando}, {Izumiura}, {Sato},
  {Hartmann}, {Hatzes}, {Barban}, {Berthomieu}, {Michel}, {Provost},
  {Turck-Chi{\`e}ze}, {Lebrun}, {Schmitt}, {Bertaux}, {Benatti}, {Claudi},
  {Cosentino}, {Leccia}, {Frandsen}, {Brogaard}, {Glowienka}, {Grundahl},
  {Stempels}, {Arentoft}, {Bazot}, {Christensen-Dalsgaard}, {Dall}, {Karoff},
  {Lundgreen-Nielsen}, {Carrier}, {Eggenberger}, {Sosnowska}, {Wittenmyer},
  {Endl}, {Metcalfe}, {Hekker}, \& {Reffert}}]{Procyon2}
{Bedding}, T.~R., {Kjeldsen}, H., {Campante}, T.~L., {et~al.} 2010, ApJ, 713,
  935

\bibitem[{{Borucki} {et~al.}(2010){Borucki}, {Koch}, {Basri}, {Batalha},
  {Brown}, {Caldwell}, {Caldwell}, {Christensen-Dalsgaard}, {Cochran},
  {DeVore}, {Dunham}, {Dupree}, {Gautier}, {Geary}, {Gilliland}, {Gould},
  {Howell}, {Jenkins}, {Kondo}, {Latham}, {Marcy}, {Meibom}, {Kjeldsen},
  {Lissauer}, {Monet}, {Morrison}, {Sasselov}, {Tarter}, {Boss}, {Brownlee},
  {Owen}, {Buzasi}, {Charbonneau}, {Doyle}, {Fortney}, {Ford}, {Holman},
  {Seager}, {Steffen}, {Welsh}, {Rowe}, {Anderson}, {Buchhave}, {Ciardi},
  {Walkowicz}, {Sherry}, {Horch}, {Isaacson}, {Everett}, {Fischer}, {Torres},
  {Johnson}, {Endl}, {MacQueen}, {Bryson}, {Dotson}, {Haas}, {Kolodziejczak},
  {Van Cleve}, {Chandrasekaran}, {Twicken}, {Quintana}, {Clarke}, {Allen},
  {Li}, {Wu}, {Tenenbaum}, {Verner}, {Bruhweiler}, {Barnes}, \&
  {Prsa}}]{Kepler}
{Borucki}, W.~J., {Koch}, D., {Basri}, G., {et~al.} 2010, Science, 327, 977

\bibitem[{{Campante}(2012)}]{CampantePhD}
{Campante}, T.~L. 2012, PhD thesis, Universidade do Porto

\bibitem[{{Campante} {et~al.}(2010){Campante}, {Karoff}, {Chaplin}, {Elsworth},
  {Handberg}, \& {Hekker}}]{Campante10}
{Campante}, T.~L., {Karoff}, C., {Chaplin}, W.~J., {et~al.} 2010, MNRAS, 408,
  542

\bibitem[{{Campante} {et~al.}(2011){Campante}, {Handberg}, {Mathur},
  {Appourchaux}, {Bedding}, {Chaplin}, {Garc{\'{\i}}a}, {Mosser}, {Benomar},
  {Bonanno}, {Corsaro}, {Fletcher}, {Gaulme}, {Hekker}, {Karoff}, {R{\'e}gulo},
  {Salabert}, {Verner}, {White}, {Houdek}, {Brand{\~a}o}, {Creevey}, {Do{\v
  g}an}, {Bazot}, {Christensen-Dalsgaard}, {Cunha}, {Elsworth}, {Huber},
  {Kjeldsen}, {Lundkvist}, {Molenda-{\.Z}akowicz}, {Monteiro}, {Stello},
  {Clarke}, {Girouard}, \& {Hall}}]{Campante11}
{Campante}, T.~L., {Handberg}, R., {Mathur}, S., {et~al.} 2011, A\&A, 534, A6

\bibitem[{{Campante} {et~al.}(2015){Campante}, {Barclay}, {Swift}, {Huber},
  {Adibekyan}, {Cochran}, {Burke}, {Isaacson}, {Quintana}, {Davies}, {Silva
  Aguirre}, {Ragozzine}, {Riddle}, {Baranec}, {Basu}, {Chaplin},
  {Christensen-Dalsgaard}, {Metcalfe}, {Bedding}, {Handberg}, {Stello},
  {Brewer}, {Hekker}, {Karoff}, {Kolbl}, {Law}, {Lundkvist}, {Miglio}, {Rowe},
  {Santos}, {Van Laerhoven}, {Arentoft}, {Elsworth}, {Fischer}, {Kawaler},
  {Kjeldsen}, {Lund}, {Marcy}, {Sousa}, {Sozzetti}, \& {White}}]{Kepler444}
{Campante}, T.~L., {Barclay}, T., {Swift}, J.~J., {et~al.} 2015, ApJ, 799, 170

\bibitem[{{Campante} {et~al.}(2016{\natexlab{a}}){Campante}, {Lund},
  {Kuszlewicz}, {Davies}, {Chaplin}, {Albrecht}, {Winn}, {Bedding}, {Benomar},
  {Bossini}, {Handberg}, {Santos}, {Van Eylen}, {Basu},
  {Christensen-Dalsgaard}, {Elsworth}, {Hekker}, {Hirano}, {Huber}, {Karoff},
  {Kjeldsen}, {Lundkvist}, {North}, {Silva Aguirre}, {Stello}, \&
  {White}}]{Campante16}
{Campante}, T.~L., {Lund}, M.~N., {Kuszlewicz}, J.~S., {et~al.}
  2016{\natexlab{a}}, ApJ, 819, 85

\bibitem[{{Campante} {et~al.}(2016{\natexlab{b}}){Campante}, {Schofield},
  {Kuszlewicz}, {Bouma}, {Chaplin}, {Huber}, {Christensen-Dalsgaard},
  {Kjeldsen}, {Bossini}, {North}, {Appourchaux}, {Latham}, {Pepper}, {Ricker},
  {Stassun}, {Vanderspek}, \& {Winn}}]{CampanteTESS}
{Campante}, T.~L., {Schofield}, M., {Kuszlewicz}, J.~S., {et~al.}
  2016{\natexlab{b}}, ApJ, 830, 138

\bibitem[{{Chaplin} {et~al.}(2014){Chaplin}, {Elsworth}, {Davies}, {Campante},
  {Handberg}, {Miglio}, \& {Basu}}]{ChaplinSuperNyquist}
{Chaplin}, W.~J., {Elsworth}, Y., {Davies}, G.~R., {et~al.} 2014, MNRAS, 445,
  946

\bibitem[{{Chaplin} \& {Miglio}(2013)}]{ChaplinMiglioReview}
{Chaplin}, W.~J., \& {Miglio}, A. 2013, ARA\&A, 51, 353

\bibitem[{{Chaplin} {et~al.}(2015){Chaplin}, {Lund}, {Handberg}, {Basu},
  {Buchhave}, {Campante}, {Davies}, {Huber}, {Latham}, {Latham}, {Serenelli},
  {Antia}, {Appourchaux}, {Ball}, {Benomar}, {Casagrande},
  {Christensen-Dalsgaard}, {Coelho}, {Creevey}, {Elsworth}, {Garc{\'{\i}}a},
  {Gaulme}, {Hekker}, {Kallinger}, {Karoff}, {Kawaler}, {Kjeldsen},
  {Lundkvist}, {Marcadon}, {Mathur}, {Miglio}, {Mosser}, {R{\'e}gulo},
  {Roxburgh}, {Silva Aguirre}, {Stello}, {Verma}, {White}, {Bedding},
  {Barclay}, {Buzasi}, {Dehuevels}, {Gizon}, {Houdek}, {Howell}, {Salabert}, \&
  {Soderblom}}]{Chaplin15}
{Chaplin}, W.~J., {Lund}, M.~N., {Handberg}, R., {et~al.} 2015, PASP, 127, 1038

\bibitem[{{Christensen-Dalsgaard}(2004)}]{JCD04}
{Christensen-Dalsgaard}, J. 2004, Sol.~Phys., 220, 137

\bibitem[{{Clarke}(2002)}]{Clarke02}
{Clarke}, D. 2002, A\&A, 386, 763

\bibitem[{{Cooley} \& {Tukey}(1965)}]{FFT}
{Cooley}, J.~W., \& {Tukey}, J.~W. 1965, Mathematics of Computing, 19, 297

\bibitem[{{Duvall} \& {Harvey}(1986)}]{DuvallHarvey86}
{Duvall}, Jr., T.~L., \& {Harvey}, J.~W. 1986, in NATO ASIC Proc.~169:
  Seismology of the Sun and the Distant Stars, ed. {D.~O.~Gough} (D.~Reidel
  Publishing Co.), 105

\bibitem[{{Garc{\'{\i}}a} \& {Ballot}(2008)}]{bdfilter}
{Garc{\'{\i}}a}, R.~A., \& {Ballot}, J. 2008, A\&A, 477, 611

\bibitem[{{Gregory}(2005)}]{GregoryBook}
{Gregory}, P.~C. 2005, {Bayesian Logical Data Analysis for the Physical
  Sciences: A Comparative Approach with `Mathematica' Support} (Cambridge
  University Press)

\bibitem[{{Handberg} \& {Campante}(2011)}]{HandCamp}
{Handberg}, R., \& {Campante}, T.~L. 2011, A\&A, 527, A56

\bibitem[{{Hastings}(1970)}]{Hastings}
{Hastings}, W.~K. 1970, Biometrika, 57, 97

\bibitem[{{Kallinger} {et~al.}(2014){Kallinger}, {De Ridder}, {Hekker},
  {Mathur}, {Mosser}, {Gruberbauer}, {Garc{\'{\i}}a}, {Karoff}, \&
  {Ballot}}]{Kallinger14}
{Kallinger}, T., {De Ridder}, J., {Hekker}, S., {et~al.} 2014, A\&A, 570, A41

\bibitem[{{Khintchine}(1934)}]{Khintchine}
{Khintchine}, A. 1934, Mathematische Annalen, 109, 604

\bibitem[{{Kjeldsen} {et~al.}(2008){Kjeldsen}, {Bedding}, {Arentoft}, {Butler},
  {Dall}, {Karoff}, {Kiss}, {Tinney}, \& {Chaplin}}]{Kjeldsen08}
{Kjeldsen}, H., {Bedding}, T.~R., {Arentoft}, T., {et~al.} 2008, ApJ, 682, 1370

\bibitem[{{Lundkvist} {et~al.}(2016){Lundkvist}, {Kjeldsen}, {Albrecht},
  {Davies}, {Basu}, {Huber}, {Justesen}, {Karoff}, {Silva Aguirre}, {Van
  Eylen}, {Vang}, {Arentoft}, {Barclay}, {Bedding}, {Campante}, {Chaplin},
  {Christensen-Dalsgaard}, {Elsworth}, {Gilliland}, {Handberg}, {Hekker},
  {Kawaler}, {Lund}, {Metcalfe}, {Miglio}, {Rowe}, {Stello}, {Tingley}, \&
  {White}}]{Lundkvist}
{Lundkvist}, M.~S., {Kjeldsen}, H., {Albrecht}, S., {et~al.} 2016, Nature
  Communications, 7, 11201

\bibitem[{{Metropolis} {et~al.}(1953){Metropolis}, {Rosenbluth}, {Rosenbluth},
  {Teller}, \& {Teller}}]{Metropolis}
{Metropolis}, N., {Rosenbluth}, A.~W., {Rosenbluth}, M.~N., {Teller}, A.~H., \&
  {Teller}, E. 1953, J.~Chem.~Phys., 21, 1087

\bibitem[{{Mosser} {et~al.}(2010){Mosser}, {Belkacem}, {Goupil}, {Miglio},
  {Morel}, {Barban}, {Baudin}, {Hekker}, {Samadi}, {De Ridder}, {Weiss},
  {Auvergne}, \& {Baglin}}]{Mosser10}
{Mosser}, B., {Belkacem}, K., {Goupil}, M.-J., {et~al.} 2010, A\&A, 517, A22

\bibitem[{{Murphy} {et~al.}(2013){Murphy}, {Shibahashi}, \&
  {Kurtz}}]{MurphySuperNyquist}
{Murphy}, S.~J., {Shibahashi}, H., \& {Kurtz}, D.~W. 2013, MNRAS, 430, 2986

\bibitem[{{Nyquist}(1928)}]{Nyquist}
{Nyquist}, H. 1928, Transactions of the AIEE, 47, 617

\bibitem[{{Parseval des Ch\^enes}(1806)}]{Parseval}
{Parseval des Ch\^enes}, M.-A. 1806, M\'emoires pr\'esent\'es \`a l'Institut
  des Sciences, Lettres et Arts, pars divers savans, et lus dans ses
  assembl\'ees. Sciences, math\'ematiques et physiques. (Savans \'etrangers.),
  1, 638

\bibitem[{{Press} \& {Rybicki}(1989)}]{LSP}
{Press}, W.~H., \& {Rybicki}, G.~B. 1989, ApJ, 338, 277

\bibitem[{{Rauer} {et~al.}(2014){Rauer}, {Catala}, {Aerts}, {Appourchaux},
  {Benz}, {Brandeker}, {Christensen-Dalsgaard}, {Deleuil}, {Gizon}, {Goupil},
  {G{\"u}del}, {Janot-Pacheco}, {Mas-Hesse}, {Pagano}, {Piotto}, {Pollacco},
  {Santos}, {Smith}, {Su{\'a}rez}, {Szab{\'o}}, {Udry}, {Adibekyan}, {Alibert},
  {Almenara}, {Amaro-Seoane}, {Eiff}, {Asplund}, {Antonello}, {Barnes},
  {Baudin}, {Belkacem}, {Bergemann}, {Bihain}, {Birch}, {Bonfils}, {Boisse},
  {Bonomo}, {Borsa}, {Brand{\~a}o}, {Brocato}, {Brun}, {Burleigh}, {Burston},
  {Cabrera}, {Cassisi}, {Chaplin}, {Charpinet}, {Chiappini}, {Church},
  {Csizmadia}, {Cunha}, {Damasso}, {Davies}, {Deeg}, {D{\'{\i}}az}, {Dreizler},
  {Dreyer}, {Eggenberger}, {Ehrenreich}, {Eigm{\"u}ller}, {Erikson}, {Farmer},
  {Feltzing}, {de Oliveira Fialho}, {Figueira}, {Forveille}, {Fridlund},
  {Garc{\'{\i}}a}, {Giommi}, {Giuffrida}, {Godolt}, {Gomes da Silva},
  {Granzer}, {Grenfell}, {Grotsch-Noels}, {G{\"u}nther}, {Haswell}, {Hatzes},
  {H{\'e}brard}, {Hekker}, {Helled}, {Heng}, {Jenkins}, {Johansen},
  {Khodachenko}, {Kislyakova}, {Kley}, {Kolb}, {Krivova}, {Kupka}, {Lammer},
  {Lanza}, {Lebreton}, {Magrin}, {Marcos-Arenal}, {Marrese}, {Marques},
  {Martins}, {Mathis}, {Mathur}, {Messina}, {Miglio}, {Montalban}, {Montalto},
  {Monteiro}, {Moradi}, {Moravveji}, {Mordasini}, {Morel}, {Mortier},
  {Nascimbeni}, {Nelson}, {Nielsen}, {Noack}, {Norton}, {Ofir}, {Oshagh},
  {Ouazzani}, {P{\'a}pics}, {Parro}, {Petit}, {Plez}, {Poretti}, {Quirrenbach},
  {Ragazzoni}, {Raimondo}, {Rainer}, {Reese}, {Redmer}, {Reffert},
  {Rojas-Ayala}, {Roxburgh}, {Salmon}, {Santerne}, {Schneider}, {Schou},
  {Schuh}, {Schunker}, {Silva-Valio}, {Silvotti}, {Skillen}, {Snellen}, {Sohl},
  {Sousa}, {Sozzetti}, {Stello}, {Strassmeier}, {{\v S}vanda}, {Szab{\'o}},
  {Tkachenko}, {Valencia}, {Van Grootel}, {Vauclair}, {Ventura}, {Wagner},
  {Walton}, {Weingrill}, {Werner}, {Wheatley}, \& {Zwintz}}]{PLATO}
{Rauer}, H., {Catala}, C., {Aerts}, C., {et~al.} 2014, Experimental Astronomy,
  38, 249

\bibitem[{{Shannon}(1949)}]{Shannon}
{Shannon}, C.~E. 1949, Proceedings of the Institute of Radio Engineers, 37, 10

\bibitem[{{Stello} {et~al.}(2009){Stello}, {Chaplin}, {Basu}, {Elsworth}, \&
  {Bedding}}]{Stello09}
{Stello}, D., {Chaplin}, W.~J., {Basu}, S., {Elsworth}, Y., \& {Bedding}, T.~R.
  2009, MNRAS, 400, L80

\bibitem[{{Stello} {et~al.}(2007){Stello}, {Bruntt}, {Kjeldsen}, {Bedding},
  {Arentoft}, {Gilliland}, {Nuspl}, {Kim}, {Kang}, {Koo}, {Lee}, {Sterken},
  {Lee}, {Jensen}, {Jacob}, {Szab{\'o}}, {Frandsen}, {Csubry}, {Dind},
  {Bouzid}, {Dall}, \& {Kiss}}]{Stello07}
{Stello}, D., {Bruntt}, H., {Kjeldsen}, H., {et~al.} 2007, MNRAS, 377, 584

\bibitem[{{Verner} {et~al.}(2011){Verner}, {Elsworth}, {Chaplin}, {Campante},
  {Corsaro}, {Gaulme}, {Hekker}, {Huber}, {Karoff}, {Mathur}, {Mosser},
  {Appourchaux}, {Ballot}, {Bedding}, {Bonanno}, {Broomhall}, {Garc{\'{\i}}a},
  {Handberg}, {New}, {Stello}, {R{\'e}gulo}, {Roxburgh}, {Salabert}, {White},
  {Caldwell}, {Christiansen}, \& {Fanelli}}]{Verner11}
{Verner}, G.~A., {Elsworth}, Y., {Chaplin}, W.~J., {et~al.} 2011, MNRAS, 415,
  3539

\bibitem[{{Wiener}(1930)}]{Wiener}
{Wiener}, N. 1930, Acta Mathematica, 55, 117

\bibitem[{{Yu} {et~al.}(2016){Yu}, {Huber}, {Bedding}, {Stello}, {Murphy},
  {Xiang}, {Bi}, \& {Li}}]{Yu16}
{Yu}, J., {Huber}, D., {Bedding}, T.~R., {et~al.} 2016, MNRAS, 463, 1297

\end{thebibliography}

\end{document}